%Paper: hep-th/9502054
%From: David Wiltshire <dlw@physics.adelaide.edu.au>
%Date: Wed, 8 Feb 1995 15:59:15 +1030 (CST)
%Date (revised): Fri, 10 Feb 1995 16:43:49 +1030 (CST)
%Date (revised): Fri, 10 Feb 1995 16:58:49 +1030 (CST)
%Date (revised): Tue, 18 Apr 1995 17:12:20 +0930 (CST)

% Small revision of 18/4/95 (earlier "revisions" just being fixing the figures
% file): in the introduction one paragraph has been added before the paragraph
% containing equation (5), and one extra footnote - no. 3 - with extra
% references [14] and [15]. To appear in Classical and Quantum Gravity.

%% This paper has seven figures (encapsulated postscript) appended in a second
%% part as a uuencoded compressed tar file with instructions for unpacking.
%% We recommend that you use epsf.tex (default), as we cannot guarantee that
%% the figures will print in the correct place on the page if you try to print
%% them separately. If you do not have epsf.tex (it comes with the dvips
%% driver) then print the figures separately and comment out the next line:
\input epsf

%% If like us your printer has the memory of a pea-brain you may find that the
%% entire postscript file (1.4 MB) is too much for the poor thing to handle
%% due to the large figure files, inefficiently produced with a scanner. In
%% that case print any offending pages, e.g. p. 17, 19, 20 (default or pp. 14,
%% 16, 17 in an A4 reduced version as below). If this exercises your patience
%% too much just print it without figures by commenting out the \input epsf
%% line above and we will be happy to supply the figures on request.

\input phyzzx
%% If use A4 sized paper - i.e. you don't live in North America - uncommenting
%% the following line will decrease the length of the paper by 4 pages:
%\vsize=10true in\voffset=-5true mm

\overfullrule=0pt \referenceminspace=36pt \hsize=6.5truein
\rightline{\vbox{\halign{#\hfil\cr ADP-95-9/M27\cr hep-th/9502054\cr}}}
\bigskip\bigskip\baselineskip=24pt
%\hsize=15truecm\hoffset=13.5true mm %% for CQG

\title{\seventeenbf Dyonic Dilaton Black Holes}\baselineskip=18pt \vfill

\author{S.J. Poletti$^1$\foot{E-mail: spoletti@physics.adelaide.edu.au},
J. Twamley$^{1,2}$\foot{E-mail: twamley@physics.uq.oz.au} and
D.L. Wiltshire$^1$\foot{E-mail: dlw@physics.adelaide.edu.au} }

\address{$^1$ Department of Physics and Mathematical Physics, University of
Adelaide,\break Adelaide, S.A. 5005, Australia.} \address{$^2$ Physics
Department, University of Queensland, St Lucia, QLD 4072, Australia.}\vfill
\def\scr{\scriptstyle}\def\Z#1{_{\lower2pt\hbox{$\scr#1$}}}\def\LA{\Lambda}
\def\g#1{{\rm g}\Z#1}\def\RN{Reissner-Nordstr\"om}

\abstract

The properties of static spherically symmetric black holes, which are both
electrically and magnetically charged, and which are coupled to the dilaton in
the presence of a cosmological constant, $\LA$, are considered. It is shown
that apart from the \RN-de Sitter solution with constant
dilaton, such solutions do not exist if $\LA>0$ (in arbitrary spacetime
dimension $\ge4$). However, asymptotically anti-de Sitter dyonic black hole
solutions with a non-trivial dilaton do exist if $\LA<0$. Both these
solutions and the asymptotically flat ($\LA=0$) solutions are studied
numerically for arbitrary values of the dilaton coupling parameter, $\g0$, in
four dimensions. The asymptotically flat solutions are found to exhibit two
horizons if $\g0=0,1,\sqrt{3},\sqrt{6},\dots,\sqrt{n(n+1)/2},\dots$, and one
horizon otherwise. For asymptotically anti-de Sitter solutions the result is
similar, but the corresponding values of $\g0$ are altered in a non-linear
fashion which depends on $\LA$ and the mass and charges of the black holes.
All dyonic solutions with $\LA\le0$ are found to have zero Hawking
temperature in the extreme limit, however, regardless of the value of $\g0$.

\vfill \centerline{(January, 1995)}
%\centerline{PACS numbers: 04.70.Bw, 04.20.Jb, 04.70.Dy, 11.25.-w} %%for CQG
\endpage \baselineskip=16pt plus.2pt minus.2pt
%\baselineskip=20pt plus.2pt minus.2pt %%for CQG
%%%%%%%%%%%%%%%%%%%%%%%%%%%%%%%%%%%%%%%%%%%%%%%%%%%%%%%%%References
\REF\GM{G.W. Gibbons and K. Maeda, \NP{B298} (1988) 741.}
\REF\GHS{D. Garfinkle, G.T. Horowitz and A. Strominger, \PR{D43} (1991) 3140;
(E) {\bf 45} (1992) 3888.}
\REF\fay{F. Dowker, J.P. Gauntlett, D.A. Kastor and J. Traschen, \PR{D49}
(1994) 2909; %hep-th/9309075
F. Dowker, J.P. Gauntlett, S.B. Giddings and G.T. Horowitz, \PR{D50} (1994)
2662.}%hep-th/9312172
\REF\HW{C.F.E. Holzhey and F. Wilczek, \NP{B380} (1992) 447.}%hep-th/9202014
%\REF\HL{B. Harms and Y. Leblanc, \PR{D46} (1992) 2334.}%hep-th/9205021
\REF\ent{G.W. Gibbons and R.E. Kallosh, Preprint NI-94-003, hep-th/9407118;\br
S.W. Hawking, G.T. Horowitz and S.F. Ross, Preprint NI-94-012, gr-qc/9409013.}
\REF\DIN{J.P. Derendinger, L.E. Ib\'a\~nez and H.P. Nilles, \PL{155B} (1985)
65; \br M. Dine, R. Rohm, N. Seiberg and E. Witten, \PL{156B} (1985) 55.}
\REF\DP{T. Damour and A.M. Polyakov, \NP{B423} (1994) 532.}%hep-th/9401069.
\REF\GH{R. Gregory and J.A. Harvey, \PR{D47} (1993) 2411.}%hep-th/9209070
\REF\HH{J.H. Horne and G.T. Horowitz, \NP{B399} (1993) 169.}%hep-th/9210012
\REF\PW{S.J. Poletti and D.L. Wiltshire, \PR{D50} (1994), 7260.}
%gr-qc/9407021
\REF\PTW{S.J. Poletti, J. Twamley and D.L. Wiltshire, Preprint ADP-94-17/M24,
hep-th/9412076, \PR{D51} (1995), in press.}
\REF\Ok{T. Okai, Preprint UT-679 (1994), hep-th/9406126.}
\REF\MW{S. Mignemi and D.L. Wiltshire, \CQG{6} (1989) 987; \PR{D46} (1992)
1475; %hep-th/9202031
D.L. Wiltshire, \PR{D44}, 1100 (1991).}
\REF\Ra{S.K. Rama, Preprint TCD-1-94 (1994), hep-th/9402009, \MPL{A}, to
appear; Preprint MRI-PHY-15-94 (1994), hep-th/9411076;\br
F. Belgiorno and A.S. Cattaneo, \IJMP{A10} (1995) 527.}
\REF\CHM{K.C.K. Chan, J.H. Horne and R.B. Mann, Preprint DAMTP-R-95/7
(1995),\break gr-qc/9402042.}
\REF\KT{D. Kastor and J. Traschen, \PR{D47} (1993) 5370.}%hep-th/9212035
\REF\HoHo{J.H. Horne and G.T. Horowitz, \PR{D48} (1993) R5457.}%hep-th/9307177
\REF\MS{T. Maki and K. Shiraishi, \CQG{10} (1993) 2171.}
\REF\axi{A. Shapere, S. Trivedi and F. Wilczek, \MPL{A6} (1991) 2677.}
\REF\DM{P. Dobiasch and D. Maison, \GRG{14} (1982) 231.}
\REF\Gi{G.W. Gibbons, \NP{B207} (1982) 337.}
\REF\GW{G.W. Gibbons and D.L. Wiltshire, \AP{N.Y.}{167} (1986) 201; (E) {\bf
176} (1987) 393.}
%%%%%%%%%%%%%%%%%%%%%%%%%%%%%%%%%%%%%%%%%%%%%%%%%%%%%%%%Shorthand
\font\sixrm=cmr6 \font\sixi=cmmi6 \font\sixsy=cmsy6 \font\sixbf=cmbx6
\font\eightrm=cmr8 \font\eighti=cmmi8 \font\eightsy=cmsy8 \font\eightbf=cmbx8
\font\eighttt=cmtt8 \font\eightit=cmti8 \font\eightsl=cmsl8 
  \font\ser=cmssi10
  
\def\tenpoint{\def\rm{\fam0\tenrm} \textfont0=\tenrm \scriptfont0=\sevenrm
\scriptscriptfont0=\fiverm \textfont1=\teni \scriptfont1=\seveni
\scriptscriptfont1=\fivei \textfont2=\tensy \scriptfont2=\sevensy
\scriptscriptfont2=\fivesy \textfont3=\tenex \scriptfont3=\tenex
\scriptscriptfont3=\tenex \textfont\itfam=\tenit \def\it{\fam\itfam\tenit}
\textfont\slfam=\tensl \def\sl{\fam\slfam\tensl} \textfont\ttfam=\tentt
\def\tt{\fam\ttfam\tentt} \textfont\bffam=\tenbf \scriptfont\bffam=\sevenbf
\scriptscriptfont\bffam=\fivebf \def\bf{\fam\bffam\tenbf}
\setbox\strutbox=\hbox{\vrule height8.5pt depth3.5pt width0pt} \let\sc=\eightrm
\let\big=\tenbig \rm} \def\eightpoint{\def\rm{\fam0\eightrm}
\textfont0=\eightrm \scriptfont0=\sixrm \scriptscriptfont0=\fiverm
\textfont1=\eighti \scriptfont1=\sixi \scriptscriptfont1=\fivei
\textfont2=\eightsy \scriptfont2=\sixsy \scriptscriptfont2=\fivesy
\textfont3=\tenex \scriptfont3=\tenex \scriptscriptfont3=\tenex
\textfont\itfam=\eightit \def\it{\fam\itfam\eightit} \textfont\slfam=\eightsl
\def\sl{\fam\slfam\eightsl} \textfont\ttfam=\eighttt
\def\tt{\fam\ttfam\eighttt} \textfont\bffam=\eightbf \scriptfont\bffam=\sixbf
\scriptscriptfont\bffam=\fivebf \def\bf{\fam\bffam\eightbf}
\setbox\strutbox=\hbox{\vrule height7pt depth2pt width0pt} \let\sc=\sixrm
\let\big=\eightbig \rm} \def\br{\hfil\break} \def\rarr{\rightarrow}
\def\scrscr{\scriptscriptstyle}  
\def\begincaption{\medskip\openup-1\jot\eightpoint}
\def\endcaption{\tenpoint\openup1\jot\leftskip=0pt\rightskip=0pt}
\def\caption#1#2{\message{#1}\begincaption\leftskip=15true mm\rightskip=15true
mm\vbox{\halign{\vtop{\parindent=0pt\parskip=0pt\strut##\strut}\cr{\bf#1}\quad
#2\cr}}\endcaption} \def\dsp{\displaystyle}
\def\X#1{_{\lower2pt\hbox{$\scrscr#1$}}}
\def\ns#1{_{\hbox{\sevenrm #1}}}
\def\ep{\epsilon}
\def\ph{\phi}\def\pt{\partial}\def\OM{\Omega}\def\dd{{\rm d}}
\def\PL#1{Phys.\ Lett.\ {\bf#1}} 
\def\AP#1#2{Ann.\ Phys.\ (#1) {\bf#2}} \def\PR#1{Phys.\ Rev.\ {\bf#1}}
\def\CQG#1{Class.\ Quantum Grav.\ {\bf#1}}\def\NP#1{Nucl.\ Phys.\ {\bf#1}}
\def\GRG#1{Gen.\ Relativ.\ Grav.\ {\bf#1}}
\def\MPL#1{Mod.\ Phys.\ Lett.\ {\bf#1}}
\def\IJMP#1{Int.\ J.\ Mod.\ Phys.\ {\bf#1}}

\def\SI{\Sigma} \def\G#1{{\rm g}\X#1} \def\V{{\cal V}} \def\Db#1{(D-#1)}
\def\Pz#1{\PH\Z{#1}}  
 
 \def\GG{\G0^{\ 2}} \def\th{\theta} \def\PH{\Phi}
\def\al{\alpha}\def\et{\eta}\def\OO#1{{\rm O}\left(#1\right)}
\def\Ve{\V\ns{exp}}\def\Vs{\V\ns{susy}}\def\ds{{\rm ds}^2}\def\kp{2\phi}
\def\Zop{\hbox{{\ser Z\hskip-.4em Z}}}
%%%%%%%%%%%%%%%%%%%%%%%%%%Modify output macros independent of phyzzx version
\sectionstyle{{}}\def\section#1{\par \ifnum\lastpenalty=30000\else\penalty-200
\vskip\sectionskip\spacecheck\sectionminspace\fi\global\advance\sectionnumber
by 1 {\protect\xdef\sectionlabel{\the\sectionstyle{\the\sectionnumber}} \wlog
{\string\section\space\sectionlabel}}\noindent{\caps\enspace\sectionlabel.~~#1}
\par\nobreak\vskip\headskip \penalty 30000 }
\def\appendix#1{\par \ifnum\lastpenalty=30000\else\penalty-200
\vskip\sectionskip\spacecheck\sectionminspace\fi%%%\global\usechapterlabeltrue
\chapterreset \xdef\chapterlabel{#1} \wlog{\string\Appendix~\chapterlabel}
\noindent{\caps\enspace Appendix}\par\nobreak\vskip\headskip \penalty 30000}
%%%%%%%%%%%%%%%%%%%%%%%%%%Option of epsf for in-text figures / default phyzzx
\ifx\epsfbox\UnDeFiNeD\message{(NO epsf.tex, FIGURES WILL BE IGNORED)}
\def\ifig#1#2#3{\FIG#1{#2}} % Figure captions will be printed at end
\else\message{(FIGURES WILL BE INCLUDED)}
\def\ifig#1#2#3{\FIGNUM#1\goodbreak\midinsert\centerline{#3}%
\smallskip\centerline{\vbox{\baselineskip12pt \advance\hsize by
-1truein\eightpoint\noindent{\bf Fig.~\the\figurecount:} #2}} %\vskip12pt
\endinsert} \def\figout{{}} \fi
%%%%%%%%%%%%%%%%%%%%%%%%%%Number footnotes
\footsymbolcount=0\def\foot{\advance\footsymbolcount by 1
\attach{\the\footsymbolcount}\Vfootnote{$^{\the\footsymbolcount}$}}
%%%%%%%%%%%%%%%%%%%%%%%%%%%%%%%%%%%%%%%%%%%%%%%%%%%%%%%%Paper body

\section{Introduction}

In effective low-energy theories of gravity derived from string theory Einstein
gravity is supplemented by additional fields such as the axion, gauge fields,
and the scalar dilaton which couples in a non-trivial way to the other fields.
The observation that black holes with non-trivial couplings to the dilaton
field have properties which sometimes differ quite dramatically from those of
the corresponding classical solutions of the standard Einstein theory
[\GM,\GHS] has provided fertile ground over the past few years for the
development of new ideas concerning quantum gravitational phenomena, such as
pair creation of black holes [\fay], black hole evaporation [\HW] and the
nature of black hole entropy [\ent].

The aim of the present paper is to extend the known dilaton black hole
solutions to include analogues of the \RN-(anti)-de Sitter metrics with both
electric and magnetic charges. In addition we will also discuss the case of
asymptotically flat dyonic solutions for values of the dilaton coupling
constant for which exact solutions are not known. In considering spacetimes
which are not asymptotically flat in the context of low-energy string inspired
gravity models the first question one should ask is what is the appropriate
``cosmological term'' one should consider. Instead of simply taking this term
to be a cosmological constant, one should conceivably take some sort of dilaton
potential. An appropriate action might be
$$\eqalign{S=\int\dd^Dx\sqrt{-g}\Biggl\{{{\cal R}\over4}-&{1\over D-2}\,g^{\mu
\nu}\pt_\mu\ph\,\pt_\nu\ph-\V(\ph)-{1\over4}\exp\left(-4\g0\ph\over D-2\right)F
_{\mu\nu}F^{\mu\nu}\cr&-{1\over2(D-2)!}\exp\left(-4\g0\ph\over D-2\right)F_{\mu
\X1\mu\X2\dots\mu\X{\!D-2}}F^{\mu\X1\mu\X2\dots\mu\X{\!D-2}}\Biggr\},\cr}\eqn
\action$$
which includes gravity, an electromagnetic field $F_{\mu\nu}$, the $\Db2$-form
field strength of an additional\foot{Our inclusion of the $\Db2$-form is not
demanded by the requirements of string theory, but is included as to allow for
generalised dyon solutions in the case of arbitrary spacetime dimension. We are
of course primarily interested in the case of four dimensions when the
$\Db2$-form coincides with the electromagnetic field.} abelian gauge field
$F_{\mu\X1\mu\X2\dots\mu\X{\!D-2}}$, and the dilaton, $\ph$, with a non-trivial
dilaton potential, $\V(\ph)$, which might realistically be expected to take a
form such as
$$\V=\Ve+\Vs\eqn\vvv$$
where
$$\Ve={\LA\over2}\exp\left(-4\g1\ph\over D-2\right)\,,\eqn\Vexp$$
and
$$\Vs={1\over4}\exp\left[-\al e^{-\kp}\right]\left\{A\Z1e^{\kp}+A\Z2+A\Z3e^{-
\kp}\right\}\,.\eqn\Vsusy$$
Here $\g0$, $\g1$, $\LA$, $\al$, $A\Z1$, $A\Z2$ and $A\Z3$ are constants.
Eqns.\ \action--\Vsusy\ are somewhat more general than is demanded by string
theory. However, if we set $\g0=1$ we obtain the standard tree-level coupling
between the dilaton and the electromagnetic field, while setting $\g1=-1$, $\LA
=\left(D_{\hbox{\sevenrm crit}}-D_{\hbox{\sevenrm eff}}\right)/(3\al')$, in the
Liouville-type term \Vexp\ yields the case of a potential corresponding to a
central charge deficit. The term \Vsusy, on the other hand, is the type of
potential which arises in four dimensions from supersymmetry breaking via
gaugino condensation in the hidden sector of the string theory\foot{The
particular potential given in \Vsusy\ is relevant for one gaugino
condensation.} [\DIN].

It is generally believed that a term such as $\Vs$ should be present since
long-range scalar forces would be present if the dilaton were massless, which
is phenomenologically unacceptable, although Damour and Polyakov have argued
that a massless but very weakly coupled dilaton can be reconciled with present
observational bounds [\DP]. If one accepts that the dilaton should be massive
then the absence of a dilaton mass term is a major physical defect with most
dilaton black hole solutions studied to date. However, Gregory and Harvey [\GH]
and Horne and Horowitz [\HH] have recently made some investigation of black
hole models which include a mass term, in the form of a simple quadratic
potential [\GH,\HH], $\V=2m^2\left(\ph-\ph\Z0\right)^2$, or alternatively of
the form [\GH] $\V=2m^2\cosh^2\left(\phi-\ph\Z0\right)$. While a rigorous proof
of the existence of black hole solutions in these models has still to be given,
the arguments of Horne and Horowitz [\HH] are nonetheless compelling.
Furthermore, it is clear from the arguments of [\GH,\HH] that the properties of
the solutions with a massive dilaton are essentially the same as those with a
massless dilaton in the case of black holes which are small with respect to the
Compton wavelength of the dilaton, which does provide some further
justification for studying the solutions with a massless dilaton.

In the present paper we will not consider potentials such as $\Vs$ or other
potentials $\V(\ph)$ which give the dilaton a mass, but will restrict our
attention to cosmological terms of the type $\Ve$. As we will see, the
solutions we will discuss here do nonetheless have some similar features to
those of [\GH,\HH], although there are also some important differences. Apart
from our own recent papers [\PW,\PTW] and that of Okai [\Ok], there has been
relatively little work on dilaton black hole solutions with Liouville-type
potentials or a simple cosmological constant\foot{Uncharged solutions in
such models [\MW,\Ra] generically either possess naked singularities or have
unusual asymptotics, or both. Amongst the class of solutions with
unusual asymptotics one can find solutions for which the curvature vanishes
asymptotically even though the solutions are not asymptotically flat according
to the conventional definition [\PW,\CHM]. It is possible that the subset of
such solutions which also possess a regular horizon could have a consistent
physical interpretation as black holes. However, we will not consider
such solutions here.}. Kastor-Traschen type [\KT]
cosmological multi-black hole solutions have been discussed by Horne and
Horowitz [\HoHo], and by Maki and Shiraishi [\MS]. However, exact solutions
have been constructed only for certain special values of the dilaton coupling
and for special powers of a Liouville-type dilaton potential [\MS]: this
included the case $\g0=1$, $\g1=-1$ relevant to a string theoretic model with a
central charge deficit, but excluded the case $\g1=0$ for which the potential
is simply a cosmological constant. Furthermore, all solutions with non-zero
dilaton couplings obtained by Maki and Shiraishi involve a dilaton which
depends on certain special powers of the time-dependent cosmic scale factor,
and thus in particular they possess no static limit.

Although a Liouville type potential \Vexp\ with $\g1=-1$ might be the most
natural choice from the point of view of string theory, such a model
unfortunately does not possess static spherically symmetric solutions with
``realistic'' asymptotics, namely solutions which are either asymptotically
flat or asymptotically of constant curvature. This conclusion is based on an
analysis of the relevant phase space for solutions which are either
electrically or magnetically charged [\PW]. It was found in [\PW] that the only
case in which ``realistic'' asymptotics are admitted, is the case $\g1=0$
corresponding to a simple cosmological constant, even allowing for the relaxed
regularity assumptions on the dilaton which are consistent with the weak
coupling limit of string theory. Although dyonic solutions were not explicitly
considered in [\PW] there is no reason to expect the conclusion to be any
different for dyons than for solutions which carry a single charge. We will
therefore henceforth simply consider the case $\g1=0$ in which the action
\action\ contains a cosmological constant.

It does not seem likely that there is any role for a pure cosmological constant
in the context of string theory. However, if a non-perturbative dilaton
potential, $\V(\ph)$, is present and if this potential has a
minimum, $\ph\Z0$, for which $\V(\ph\Z0)\ne0$, then asymptotic regions for
which $\ph\rarr\ph\Z0$ would correspond to a background universe with a
non-zero (hopefully small) vacuum energy. In the case of certain potentials
with a minimum for which $\V(\ph\Z0)=0$ it has been shown [\GH,\HH] that the
appropriate dilaton black hole solutions are well approximated by the solutions
with a massless dilaton in various regimes. We would hope that the solutions
which we discuss here would similarly provide a useful approximation for
corresponding models in which $\V(\ph\Z0)\ne0$. Thus although a pure
cosmological constant is not favoured by string theory, we believe that the
models investigated here could nevertheless be of considerable physical
interest as a limiting case of more physically motivated stringy black
hole solutions.

The properties of dilaton black holes in the presence of a cosmological
constant which carry either an electric or a magnetic charge were dealt with in
our recent paper [\PTW]. Here we will consider the extension of this work to
solutions which carry both charges. To be specific, let us assume that the
spacetime metric is static and spherically symmetric, taking the form
$$\ds=-f\dd t^2+f^{-1}\dd r^2+R^2\dd\OM^2\Z{D-2},\eqn\coorda$$
where $f=f(r)$ and $R=R(r)$, and $\dd\OM^2\Z{D-2}$ is the standard round metric
on a $\Db2$-sphere, with angular coordinates $\th_i$, $i=1\dots D-2$. The
electromagnetic field and the $\Db2$-form field will then be assumed to have
components given in an orthonormal frame by
$$F_{\hat t\hat r}=-F_{\hat r\hat t}=\exp\left(4\g0\ph\over D-2\right){Q\over R
^{D-2}}\,,\eqn\elec$$
and
$$F_{\hat\th\X1\hat\th\X2\dots\hat\th\X{\!D-2}}={P\over R^{D-2}}\ep_{\hat\th\X1
\hat\th\X2\dots\hat\th\X{\!D-2}}\eqn\mag$$
respectively. The ansatz \elec\ is appropriate to an isolated electric charge,
and in four dimensions \mag\ becomes the ansatz for an isolated magnetic
monopole, and thus with both fields present we obtain a generalised dyon
ansatz\foot{In the context of dilaton gravity the term {\it dyon} is sometimes
applied to solutions with both electric (or magnetic) and {\it axionic} charges
[\axi]. We are not considering axionic dyons here.}.

\section{Non-existence of non-trivial dyonic dilaton black holes with a
positive cosmological constant}

Let us firstly consider the case of a positive cosmological constant $\V\equiv
\LA/2$, with $\LA>0$. It is quite straightforward to show that apart from one
trivial case dyonic black holes with a positive cosmological constant do not
exist in the presence of the dilaton field. For this purpose it is convenient
to define a constant
$$a={2\g0\over\Db2}\,,\eqn\aaa$$
and to rescale the dilaton by adding a constant
$$\PH\equiv\ph+{1\over2a}\ln\left|Q\over P\right|.\eqn\scaledil$$
The field equations derived from \action\ may then be written compactly as
$$\eqalignno{{1\over R^{D-2}}\left[R^{D-2}f\PH'\right]'=\;&{a\Db2|QP|\sinh(2a
\PH)\over R^{2\Db2}}\,,&\eqname\feaA\cr{R''\over R}=\;&-{4\PH'^2\over\Db2^2}&
\eqname\feaB\cr{1\over R^{D-2}}\left[f\left(R^{D-2}\right )'\right]'=\;&\Db2\Db
3{1\over R^2}-2\LA-{4|QP|\cosh(2a\PH)\over R^{2\Db2}}\,,&\eqname\feaC\cr{1\over
R^{2D-6}}\left[fR^{2D-6}\right]''-&\,\Db4{R'\over R^{2D-5}}\left[fR^{2D-6}
\right]'={2\Db3^2\over R^2}-4\LA,&\eqname\feaD\cr}$$
with $'\equiv d/dr$. Eqn.\ \feaD\ follows from \feaA--\feaC\ by virtue of the
Bianchi identity if $\PH'\ne0$.

It is possible to have asymptotically flat solutions to these field equations
if $\LA=0$, while asymptotically de Sitter ($\LA>0$) or asymptotically anti-de
Sitter ($\LA<0$) solutions are obtained otherwise. Furthermore, de Sitter
asymptotics are obtained only in the region in which the Killing vector $\pt/
\pt t$ is spacelike, and consequently any black hole solutions in such a model
must possess at least two horizons. It is quite straightforward to show that in
fact there are no such solutions. We prove the result by contradiction.

Suppose that black hole solutions do exist with at least two regular horizons,
and let the two outermost horizons be labelled $r_\pm$, with $r_-<r_+$. The
requirement of regularity at the horizon means that near $r=r_+$, $f\propto(r-r
_+)$ and $\ph(r_+)$ and $R(r_+)$ are bounded with $R(r_+)\ne0$, and similarly
for $r_-$. We may multiply \feaA\ by $\PH$ to obtain
$$\left[R^{D-2}f\PH\PH'\right]'=R^{D-2}f\PH'^2+\Db2|QP|\,{a\PH\sinh(2a\PH)\over
R^{D-2}}\,.\eqn\horz$$
If we integrate \horz\ with respect to $r$ between the two horizons, $r_-$ and
$r_+$, then the integral of the l.h.s.\ vansishes. Provided that the Killing
vector $\pt/\pt t$ is timelike in the region between the two horizons -- which
must be the case for solutions with de Sitter-like asymptotics ($\LA>0$) --
then $f(r)>0$ on the interval $(r_-,r_+)$ and consequently the r.h.s.\ of
\horz\ is positive-definite. A contradiction is thus obtained if $\LA>0$ in all
cases apart from that of a constant dilaton field with $\PH=0$, or equivalently
$e^{2a\ph}=|P|/|Q|$. To retrieve the standard normalisation for the
electromagnetic field we require that $\ph\rarr0$ at spatial infinity, and thus
$|Q|=|P|$ in the case of the solutions with a trivial constant dilaton.
Furthermore, in the trivial case eqn.\ \feaA\ is satisfied identically and the
remaining equations reduce to those obtained in the absence of a dilaton, and
thus the spacetime geometry is simply that of the standard \RN-(anti)-de Sitter
solution
$$\eqalign{R(r)&=r,\cr f(r)&={-2\LA r^2\over\Db1\Db2}+1-{2M\over r^{D-3}}+{2
\left(Q^2+P^2\right)\over\Db2\Db3r^{2D-6}}\,,}\eqn\rnds$$
with equal electric and magnetic charges, $|Q|=|P|$.

In addition, there are also Robinson-Bertotti-type solutions of the form
$\PH=0$, $R=R\ns{ext}$ and
$$f=\left[\Db3^2R\ns{ext}^{-2}-4\LA\right]+c\Z1r+c\Z2\eqn\rbA$$
where $c\Z1$ and $c\Z2$ are arbitrary constants, and the constant $R\ns{ext}$
is a solution of the polynomial
$$-2\LA R\ns{ext}^{2D-4}+\Db2\Db3R\ns{ext}^{2D-6}-4|QP|=0.\eqn\rbB$$

The argument we have given above for the non-existence of $\LA>0$ black holes
with a non-trivial dilaton of course only applies to {\it dyonic} solutions, as
the transformation \scaledil\ used in writing the field equations in the form
\feaA--\feaC\ becomes singular if either $Q=0$ or $P=0$. In the case of
solutions which carry only an electric charge or a magnetic charge one finds
that black hole solutions can have at most one horizon, for either sign of
$\LA$, as was shown in [\PTW]. Since asymptotically de Sitter black holes must
have at least two horizons, one immediately has the result that no singly
charged dilaton black holes exist if $\LA>0$. In this case even a trivial
(constant) dilaton solution is not admitted. Solutions with a single horizon do
exist, of course -- however, the horizon is a cosmological one and the
solutions possess a naked singularity as viewed by observers in the region
where $\pt/\pt t$ is timelike.

\section{Asymptotically flat dyonic dilaton black holes}

Before considering solutions in the presence of a cosmological constant, we
will begin by discussing the corresponding asymptotically flat ($\LA=0$) dyonic
dilaton black holes, so as to be able to make some comparison to the solutions
with $\LA<0$. Exact dyonic solutions are only known in the cases $\g0=\sqrt{3
\Db3}$ [\GM,\DM--\GW]\foot{In fact, Gibbons and Maeda [\GM] leave the case $\g0
=\sqrt{3\Db3}$ as an ``exercise for the reader''. For reference we list the
solution to this exercise in the Appendix.} and $\g0=\sqrt{D-3}$ [\GM,\Gi]. In
four dimensions the solutions for $\g0=1,\sqrt{3}$ may be written\foot{Our
choice of the origin of $r$ agrees with [\GW], but differs from Gibbons and
Maeda [\GM], whose corresponding radial coordinate is obtained by $r\rarr r+M$
above. The signs of $\ph$ and $\SI$ in [\GW] are opposite to those here.}
$$\eqalignno{R(r)&\;=\left[A(r)B(r)\right]^{1/(\GG+1)},&\eqname\RAB\cr f(r)&\;=
{(r-M)^2-%r\Z0^{\ 2}
\left[M^2+\SI^2-Q^2-P^2\right]\over R^2},&\eqname\FAB\cr\exp\left(2\g0\PH
\right)&\;=\left|Q\over P\right|\left[A(r)\over B(r)\right]^{2\GG/(\GG+1)}&
\eqname\EAB\cr}$$
where for $\g0=1$ (appropriate to string theory)
$$\eqalign{A(r)&=r+\SI,\cr B(r)&=r-\SI,\cr}\eqn\solA$$
$$2M\SI=P^2-Q^2,\eqn\conA$$
while for $\g0=\sqrt{3}$ (appropriate to 5-dimensional Kaluza-Klein theory)
$$\eqalign{A(r)&=\left(r-r\Z{A-}\right)\left(r-r\Z{A+}\right),\qquad r\Z{A\pm}=
-\SI/\sqrt{3}\pm\left[2P^2\SI\over\SI+\sqrt{3}M\right]^{1/2},\cr B(r)&=\left(r
-r\Z{B-}\right)\left(r-r\Z{B+}\right),\qquad r\Z{B\pm}=\SI/\sqrt{3}\pm\left[2Q^
2\SI\over\SI-\sqrt{3}M\right]^{1/2},\cr}\eqn\solB$$
$${Q^2\over\SI-\sqrt{3}M}+{P^2\over\SI+\sqrt{3}M}={2\over3}\SI.\eqn\conB$$
In both cases the scalar charge, $\SI$, is seen to depend on the other charges,
$M$, $Q$ and $P$, of the theory through relations \conA\ and \conB, and thus
does not constitute an independent black hole ``hair''.

Both solutions have the property of having two horizons if both the electric
and magnetic charges are non-zero, and thus are qualitatively much like the
ordinary \RN\ solution. In particular, the extreme solutions have an event
horizon of finite area, and also zero Hawking temperature. If one of the
charges is zero, however, then there exists only a single horizon and in the
extreme limit the area of the event horizon pinches off to zero, while the
temperature is finite for the $\g0=1$ solution [\GM], and infinite for the $\g0
=\sqrt{3}$ solution [\GW]. The fact that the temperature is formally infinite
in the second case merely signals the breakdown of the semiclassical limit if
one is considering the Hawking evaporation process. Holzhey and Wilczek [\HW]
have in fact demonstrated that in the case of the $\g0>1$ solutions an infinite
mass gap develops for quanta with a mass less than that of the black hole so
that the Hawking radiation slows down and comes to an end at the extremal
limit, despite the infinite temperature.

It is worthwhile commenting on the properties of asymptotically flat dyonic
solutions for values of $\g0$ other than $1$ or $\sqrt{3}$. Some aspects of the
global properties of the solutions for arbitrary $\g0$ and $D$ can be derived
by studying the phase space using an approach similar to that of [\PW,\MW].
However, we will simply study the solutions numerically here, as we will adopt
an identical approach for the $\LA<0$ solutions. The behaviour of the solutions
at large $r$ can be determined by making the expansions $\PH=\sum_{i=0}^\infty
\Pz ir^{-i}$, $f=\sum_{i=0}^\infty f\Z ir^{-i}$, $R=r+\sum_{i=0}^\infty R\Z
ir^{-i}$, and substituting in the field equations \feaA--\feaC\ with $\LA=0$.
If we use the freedom of translating the origin in the radial direction to set
$R\Z0=0$, which accords with the choice made in \FAB, \solA, \solB, we find
$$\eqalign{\PH=\;&\Pz0+{\SI\over r^{D-3}}+\left[M\SI+{a\Db2|QP|\sinh(2a\Pz0)
\over2\Db3^2}\right]{1\over r^{2D-6}}+\OO{1\over r^{2D-5}},\cr f=\;&1-{2M\over
r^{D-3}}+{4|QP|\cosh(2a\Pz0)\over{(D-2)(D-3)r^{2D-6}}}+\OO{1\over r^{2D-5}},\cr
R=\;&r-{2\Db3\SI^2\over{(2D-7)(D-2)^2r^{2D-7}}}+\OO{1\over r^{2D-6}}.\cr}\eqn
\assaf$$
The constants $M$, $\Pz0$ and $\SI$ are free, $M$ being proportional to the ADM
mass. We can use the freedom of adding a constant to the dilaton to set
$$\Pz0={1\over2a}\ln\left|Q\over P\right|,\eqn\setphiz$$
so that $\ph=\SI/r^{D-3}+\OO{r^{-(2D-6)}}$ by \scaledil, which yields the
standard normalisation for the electromagnetic field, and accords with the
choice made in the exact solutions above.

Although the scalar charge $\SI$ is a free parameter as far as the aymptotic
series is concerned, if we further demand that a particular solution with an
asymptotic expansion \assaf\ corresponds to a black hole, then we can integrate
equation \feaA\ between the outermost horizon, $r_+$, and infinity to obtain an
integral relation
$$\SI={-a\Db2|QP|\over2\Db3}\int_{r_+}^\infty\dd r\,{\sinh(2a\PH)\over R^{D-2}}
\,.\eqn\intrelA$$
Consequently, for black hole solutions $\SI$ is constrained to depend on the
other charges of the theory, which is of course what we would have anticipated
from the exact solutions. The sign of the scalar charge is fixed to be $\SI<0$
if $|Q|>|P|$ (electrically dominated solutions), or $\SI>0$ if $|Q|<|P|$
(magnetically dominated solutions). To see this one may observe that since the
r.h.s.\ of \horz\ is positive definite on the interval $(r_+,\infty)$, if one
integrates \horz\ from the outermost horizon $r_+$ to an arbitrary point $r_o$
in the domain of outer communications then $\PH(r_o)\PH'(r_o)>0$ if the dilaton
field is non-trivial. Thus either: (i) $\PH>0$ and $\PH$ is monotonically
increasing on the interval $(r_+,\infty)$; or (ii) $\PH<0$ and $\PH$ is
monotonically decreasing on the interval $(r_+,\infty)$. The asymptotic
expansions \assaf\ then fix the sign of $\SI$ as above, and by \setphiz\ cases
(i) and (ii) are seen to correspond to electrically- and magnetically-dominated
solutions respectively. This argument will not be changed if $\LA<0$.

It is possible to derive a further constraint which must be satisfied by the
charges in the case of the extreme solutions. In particular, if we multiply
\feaA\ by $2R^{2(D-2)}\PH'f/\Db2$ we obtain
$${1\over D-2}\left[R^{2(D-2)}f^2\PH'^2\right]'=2a|QP|\PH'f\sinh(2a\PH),\eqn
\junka$$
while the difference of $2\Db3$ times \feaC\ and $\Db2$ times \feaD\ may be
multiplied by $\Db2R^{2(D-2)}f'/\left[8\Db3\right]$ to give
$${1\over16}\left(D-2\over D-3\right)\left[R^{2(D-2)}f'^2\right]'=|QP|f'\cosh
(2a\PH).\eqn\junkb$$
If we add \junka\ to \junkb\ the resulting equation may be directly integrated
to give
$$\left[{f^2\PH'^2\over D-2}+{\Db2f'^2\over16\Db3}\right]R^{2(D-2)}=|QP|f\cosh(
2a\PH)+c\Z0\eqn\junkc$$
where $c\Z0$ is an arbitrary constant. If we evaluate the integral between
spatial infinity and a degenerate horizon at which $f=0$ and $f'=0$, and use
the asymptotic series \assaf\ we find that
$$M^2+{4\Db3\SI^2\over\Db2^2}={2\left(Q^2+P^2\right)\over\Db2\Db3^3}\eqn\relext
$$
in the extreme limit.

For the purpose of numerical integration we will follow Horne and Horowitz
[\HH] and make a change of coordinates to use $R$ as the radial variable, so
that the metric becomes
$$\ds=-f\dd t^2+h^{-1}\dd R^2+R^2\dd\OM\Z{D-2}\eqn\coorb$$
where $h(R)\equiv f\left(\dd R\over\dd r\right)^2$, and now $f=f(R)$. The
advantage of working with these coordinates is that by suitably combining the
appropriate differential equations one can solve for $f$ in terms of $h$ and
$\ph$. One finds
$$f=h\exp\left[{8\over\Db2^2}\int^R\dd\bar R\,\dot\ph^2\bar R\right]\,,\eqn
\gabi$$
where $.\equiv\dd/\dd R$. There are then just two independent field equations
remaining, viz.\
$$\eqalignno{-&R\dot h+\Db3\left(1-h\right)={4R^2h\dot\ph^2\over\Db2^2}+{2\LA R
^2\over D-2}+{4|QP|\cosh(2a\PH)\over\Db2R^{2D-6}},&\eqname\febA\cr&Rh\ddot\PH+s
(h,\PH,\dot\PH)=0,&\eqname\febB\cr}$$
where
$$s(h,\PH,\dot\PH)\equiv\dot\PH\left[D-3+h-{2\LA R^2\over D-2}-{4|QP|\cosh(2a
\PH)\over\Db2R^{2D-6}}\right]-\Db2a|QP|{\sinh(2a\PH)\over R^{2D-5}}\eqn\shoot$$
Although we are interested in solutions with $\LA=0$ at present, we have left $
\LA$ explicitly in \febA--\shoot, as the same equations will be used in the
next section with $\LA<0$. The asymptotic series \assaf\ become
$$\eqalign{\PH=\;&{1\over2a}\ln\left|Q\over P\right|+{\SI\over R^{D-3}}+\OO{1
\over R^{D-2}}\cr h=\;&1-{2M\over R^{D-3}}+{2\left(Q^2+P^2\right)\over{(D-2)(D-
3)R^{2D-6}}}+\OO{1\over R^{2D-5}}\cr}\eqn\assif$$
in terms of the new variables. For numerical integration we will fix $D=4$.

Equations \febA, \febB\ are equivalent to three first order ordinary
differential equations and thus generally have a three parameter set of
solutions. However, many of these will correspond to naked singularites. The
requirement that solutions have at least one regular horizon reduces the three
parameters to two, which may be taken to be the radial position of the
outermost horizon, $R_+$, and $\PH_+\equiv\PH(R_+)$, if we treat the
integration as an intial value problem. Since the equations are singular on the
horizon, we start the integration a small distance from $R_+$, the initial
values of $h$, $\PH$ and $\dot\PH$ being determined in terms of $R_+$ and $\PH_
+$ by solving for the coefficients $\tilde h_i$ and $\tilde\PH_i$ in the power
series expansions,
$$\eqalign{h=\sum_{i=1}^\infty\tilde h_i\left(R-R_+\right)^i,\cr\PH=\PH_++\sum_
{i=1}^\infty\tilde\PH_i\left(R-R_+\right)^i.\cr}\eqn\horexp$$
Since the solutions are rather cumbersome we will not list them here.

{}From \S2 it follows that if there is more than one regular horizon then $h<0$
between any two horizons for solutions with a non-trivial dilaton, and thus
black hole solutions with $\LA\le0$ can have at most two horizons. Since two
horizons are possible we have a choice of starting on an inner horizon, $R_-$,
with inital data $\dot h(R_-)<0$, or an outer horizon, $R_+$, with initial data
$\dot h(R_+)>0$. From the discussion above it follows that initial data with $
\PH(R_+)>0$ and $\dot\PH(R_+)>0$ will yield electrically dominated solutions,
and initial data with $\PH(R_+)<0$ and $\dot\PH(R_+)<0$ magnetically dominated
solutions.

Although the exact solutions \RAB--\solB\ both have two horizons there is no
guarantee that black holes possess two horizons for other values of $\g0$. It
is not straightforward to test for the existence of double horizon solutions if
one uses an initial value integration method, as we have chosen to do, since
the integration routine is usually halted when $h$ approaches zero. The problem
occurs since the numerical routine determines $\ddot\PH$ by setting $\ddot\Phi
=s/(h R)$, where the quantity $s$ is defined by \shoot. The numerical
integration is also singular near $h=0$ for the cases $\g0=1,\sqrt{3}$, when
the appropriate exact solutions are known to exist. To distinguish a true
second horizon therefore, we need to test whether $\ddot\PH\rarr0$, or
equivalently whether $s\rarr0$ sufficiently quickly, as the second horizon is
approached, either integrating outwards or inwards. Given that the inner
horizon often occurs at very small values of $R$, where $\dot h$ is large, to
reduce numerical errors it is convenient to set initial data at a regular inner
horizon and to integrate outwards. Having found a second horizon for which
$s(R_+)=0$, we must also check that the solution is indeed a black hole by
resuming the integration just beyond the second horizon and checking that it
does eventually match the series \assif.

As a result of such an analysis we have found that a second horizon is obtained
only if $\g0$ takes on certain discrete values. In fact, our numerical results
are consistent with $\g0$ being the square root of a triangle number:
$$\g0=0,1,\sqrt{3},\sqrt{6},\sqrt{10},\dots,\sqrt{n(n+1)/2},\dots\eqn\series$$
For other values of $\g0$ the dyonic black hole solutions only have one
horizon. As $\g0$ is varied the following pattern is observed for initial data
with $\PH(R_-)<0$: for $0<\g0<1$ the function $s(R)$ decreases from zero to a
local minimum, and then increases monotonically, crossing zero a second time
before the second zero of $h(R)$ is reached. For $\g0=1$ the pattern is the
same but the zero of $s(R)$ coincides with the second zero of $h(R)$. For $1<\g
0<\sqrt{3}$ $s(R)$ has both a local minimum and a local maximum, and for values
for which the local maximum of $s(R)$ is positive there is one additional zero
before $h(R)$ reaches its second zero. For $\g0=\sqrt{3}$ the third zero of $s
(R)$ coincides with the second zero of $h(R)$. For $\sqrt{3}<\g0<\sqrt{6}$ the
number of turning points and possible zeros of $s(R)$ increases by one, and
again the final zero of $s(R)$ coincides with the second zero of $h(R)$ only if
$\g0=\sqrt{6}$. As $\g0$ increases further the pattern is repeated with $s(R)$
oscillating between $R_-$ and $R_+$, and the number of oscillations increasing
by one every time $\g0$ attains a value in the series $\sqrt{n(n+1)/2}$,
$n\in\Zop^{\,\scrscr+}$. The behaviour of $s(R)$ is displayed in Fig.\ 1 for
the lowest critical values of $\g0$, and in Fig.\ 2 for some non-critical
values. If we choose initial data with $\PH(R_-)>0$, then similar results are
obtained since the equations are invariant under the duality transformation $Q
\leftrightarrow P$, $\PH\rarr-\PH$.\ifig\shoota{The behaviour of $R^3s$, where
$s$ is defined by \shoot\ with $\LA=0$, as a function of $R$, between regular
horizons for critical values of the dilaton coupling parameter
$\g0$.}{\epsfbox{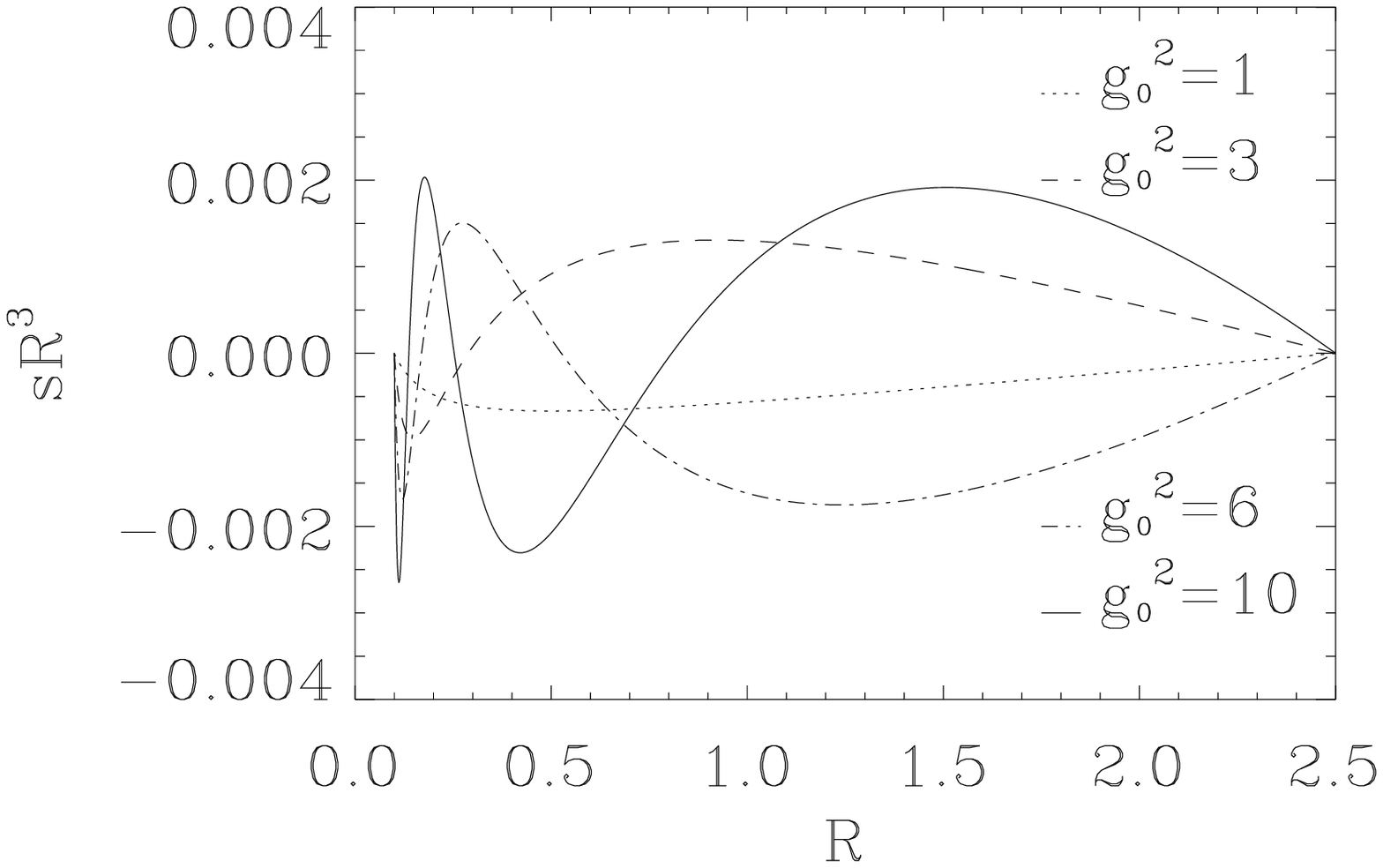}} \ifig\shootb{The behaviour of $R^3s$, where $s$
is defined by \shoot\ with $\LA=0$, as a function of $R$, in some typical cases
when the second zero of $h$ does not correspond to a regular
horizon.}{\epsfbox{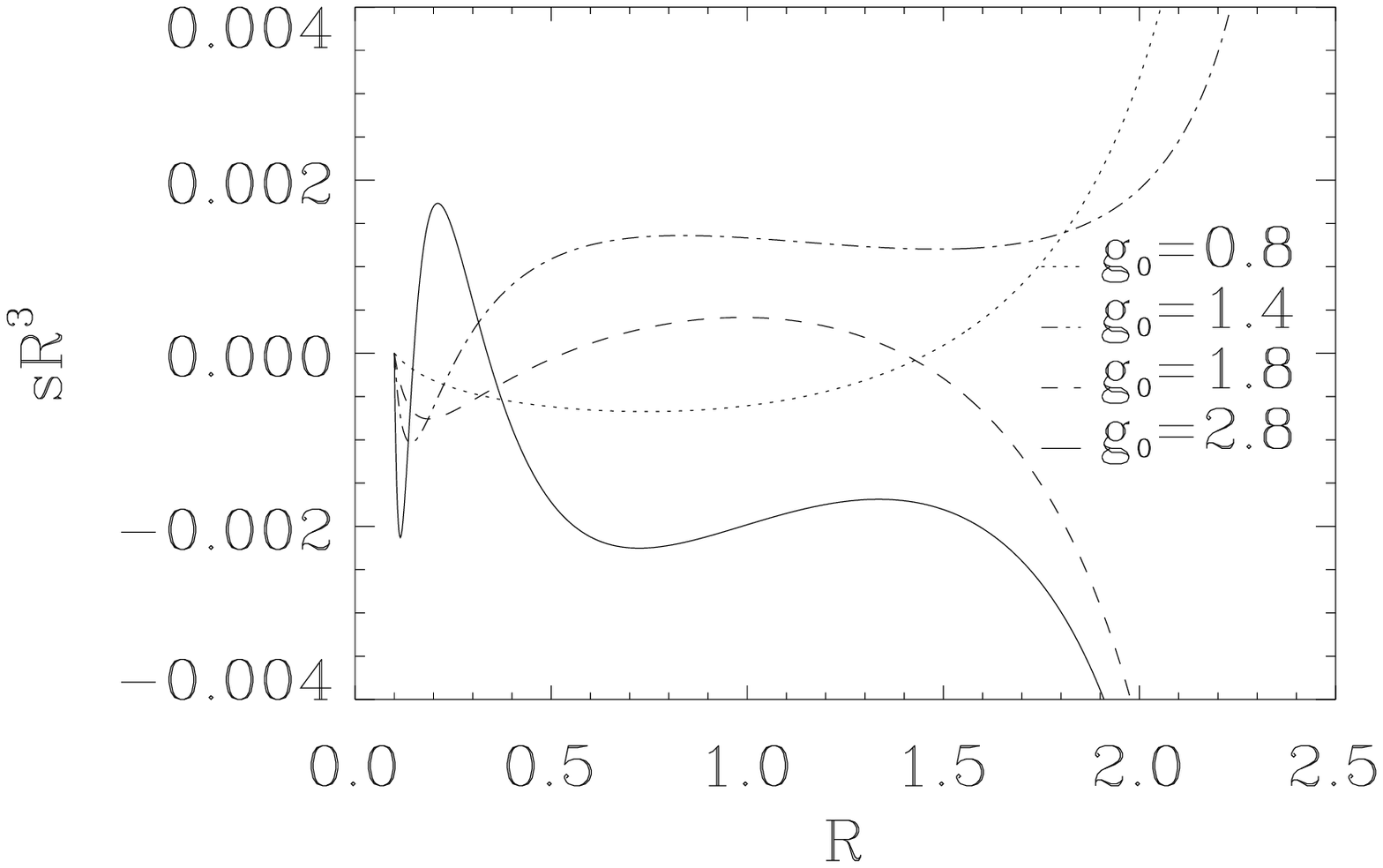}}

The critical values of $\g0$ are independent of $|QP|$ and the initial values
$R_-$ and $\PH(R_-)$, with the proviso that for each $R_-$ there is an upper
bound for $|\PH(R_-)|$, beyond which solutions are not asymptotically flat
although two regular horizons exist. In the case of larger values of $\g0$,
however, we find that the precise numerical value of $\g0$ begins to show some
dependence on the initial data for large values of $\PH(R_-)$. As the problem
increases with increasing $\g0$ it would appear to be merely the result of
numerical errors which increase rapidly for large $\g0$ due to the exponential
dependence on $\g0\PH$. Even for large $\g0$ the largest values of $\PH(R_-)$
consistent with asymptotically flat solutions yield numerical values of $\g0$
which agree with the series \series\ to $0.1\%$. We have numerically checked
terms up to $\g0=\sqrt{21}$ in the series. Since zeros of $s(R)$ correspond to
points with $\ddot\PH=0$ we also find that $\PH(R)$ oscillates between the two
horizons for solutions with critical values of $\g0$. This is illustrated in
Fig.\ 3. \ifig\phishoot{The behaviour of $\PH$ for $\LA=0$, as a function of $R
$ between the horizons $R_-$ and $R_+$, in cases that two regular horizons
exist.} {\epsfbox{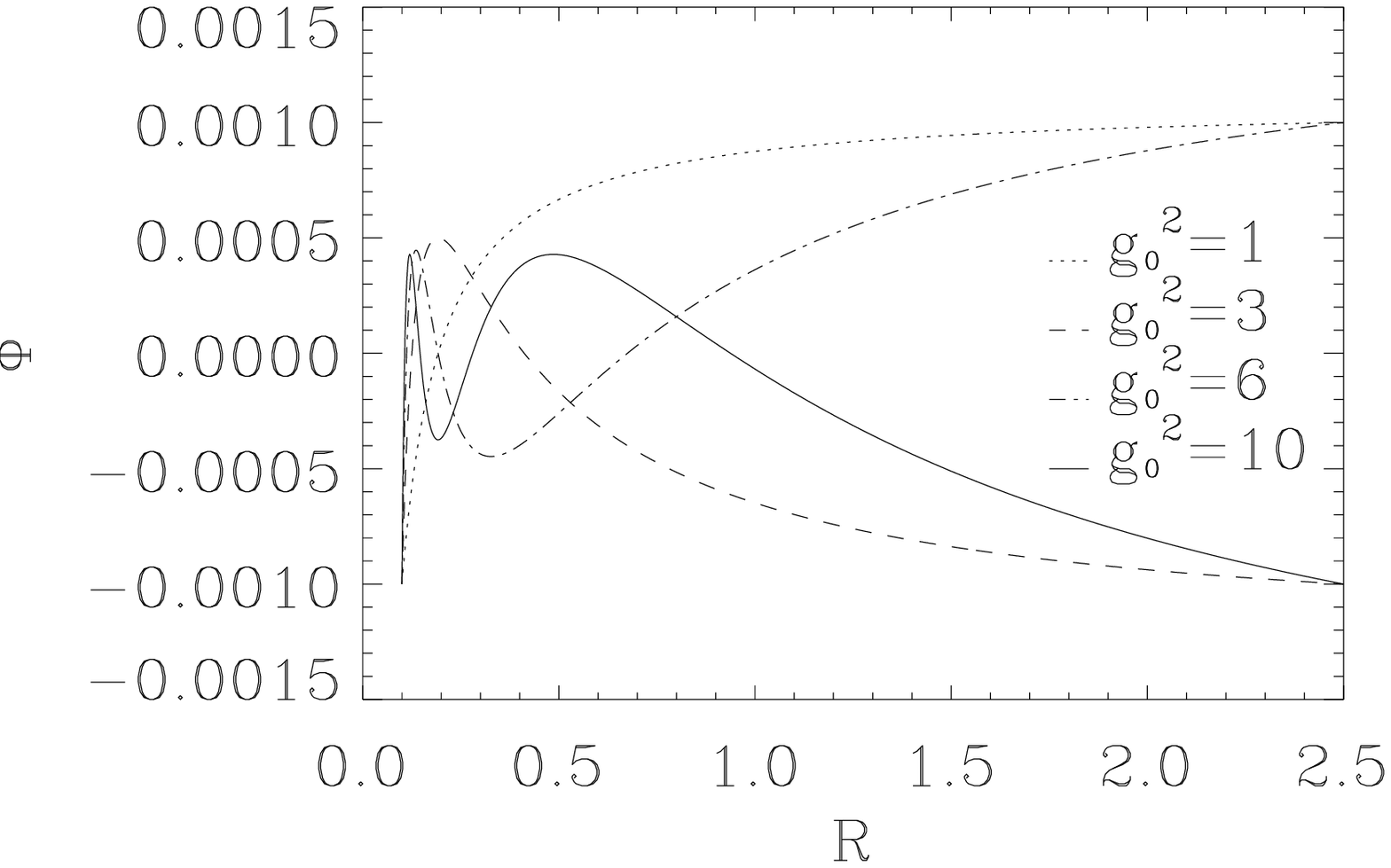}}

If $\g0$ is not equal to one of the critical eigenvalues then black holes
solutions do exist, but they have only one horizon. Physically the spacetime
geometry is no different to that of the solutions with two horizons as far as
observers in the domain of outer communications are concerned. In particular,
unlike the corresponding solutions with a single charge [\GM,\GHS,\Gi--\GW] the
extreme limit occurs for finite $R$, namely at $R\ns{ext}=\sqrt{2|QP|}$, rather
than at $R=0$. What appears to be occuring in the general case is that the
function $h(R)$ has two zeros which would both correspond to regular horizons
if it were not for the fact that in general the dilaton becomes singular at one
of the potential horizons. Only for the critical values of $\g0$ is the dilaton
regular at both zeros of $h$. In the case that the zero of $h$ is degenerate,
i.e., $h=0$ and $\dot h=0$, the dilaton is always regular at the horizon,
however. The spacetime geometry of the extreme solutions approaches that of the
Robinson-Bertotti-type solution \rbA, \rbB\ (with $\LA=0$) in the neighbourhood
of the degenerate horizon.

Since the horizon of the extreme solutions occurs at a finite value of $R$ it
follows that the entropy -- at least, the entropy na\"{\i}vely defined as a
quarter of the area of the event horizon -- is finite, while the temperature,
which is given in general by $T=\dot h(R_+)/(4\pi)$, is zero in the extreme
limit. Contour plots of isotherms are plotted in Fig.\ 4 for one non-critical
value of $\g0$ to illustrate this point. They are qualitatively similar to the
isotherms of the standard \RN\ solution. \ifig\afiso{Contours of constant
temperature, $T$, for asymptotically flat black holes ($\LA=0$), in the case
$\g0=0.5$: (a) $|P|=0.5|Q|$; (b) $|P|=2|Q|$.} {\epsfbox{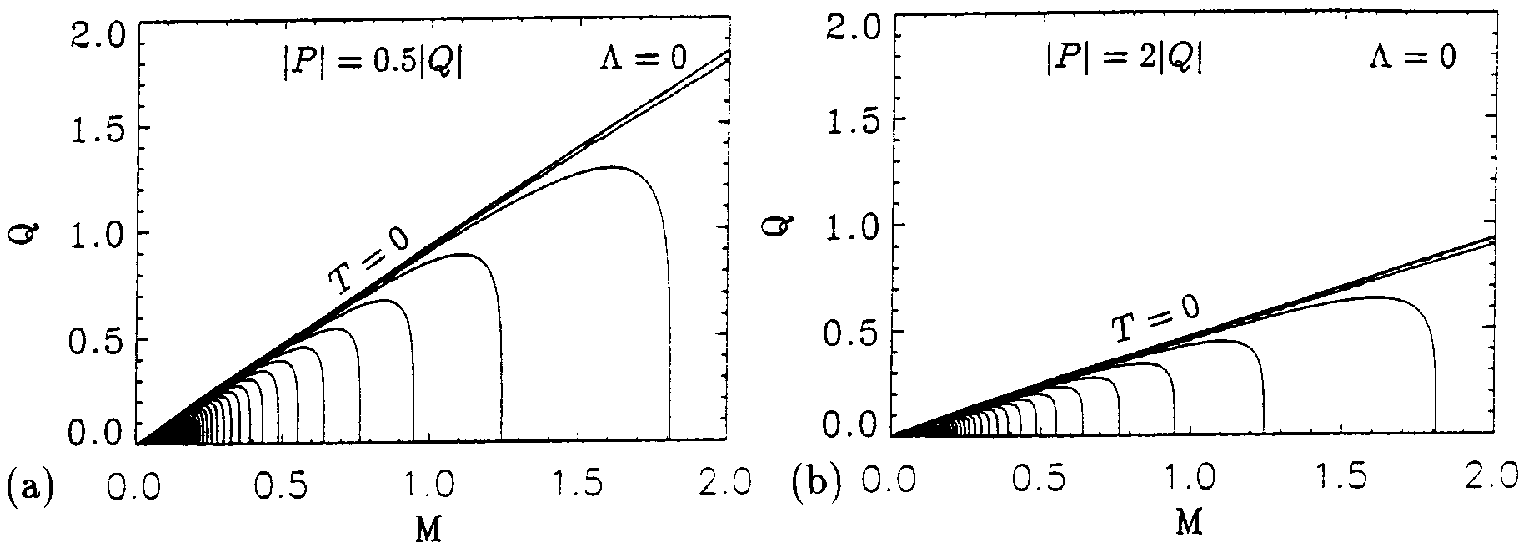}}

These contour plots are based on a large number ($\sim5\times10^3$) of
numerical integrations. The integration procedure we adopted was to fix $|QP|$
and to determine regular initial data $R_+,\PH_+$ such that $\dot h(R_+)>0$ at
a regular horizon, using solutions obtain from substituting the expansions
\horexp\ into the field equations. We then began integrations a short distance
outside $R_+$ and integrated outwards until solutions matched the asymptotic
series \assif\ to some specified accuracy -- five significant figures in our
case. Not all initial values of $R_+$ and $\PH_+$ yield asymptotically flat
solutions. However, for those that do we are able to identify $\Pz0$ according
to \setphiz. In order to fix the value of $|Q|/|P|$ so as to obtain plots such
as those shown in Fig.\ 4, we then repeated the integration using a
Newton-Raphson algorithm to vary $\PH_+$ for fixed $R_+$ until $\Pz0$ agreed
with the chosen value of $|Q|/|P|$ to some specified accuracy.

Although the extreme limit appears to be approximately linear over the range
over the range of values shown in Fig.\ 4, this is not strictly the case. In
general the relation between $M$, $P$ and $Q$ which defines the extreme
solutions is a very complicated one. This can already be seen in the case of
the $\g0=\sqrt{3}$ exact solutions, where the extreme limit is given by
$M^2+\SI^2=Q^2+P^2$, with $\SI$ being a solution of the cubic equation \conB.
Only for $\g0=1$ does the extreme limit correspond to a linear relationship
between $M$, $P$ and $Q$, namely
$$\sqrt{2}M=|P|+|Q|.\eqn\extrA$$
It is nevertheless possible to approximate the curve of extreme solutions in
the limit that the asymptotic series \assif\ are valid up to the degenerate
horizon. For example, for $D=4$ we find from the first three terms in the
series
$${\SI\over M}\approx{1\over1+\sqrt{2qp}}\left[{qp\over\g0}\ln\left|q\over p
\right|+{\g0\over2}\left(p^2-q^2\right)\right]\eqn\degsig$$
where $q\equiv Q/M$ and $p\equiv P/M$. If we substitute \degsig\ into \relext\
and solve for $q$ the result agrees with the gradient of the curve for the
largest values of $M$ found numerically in Figs.\ 4 to within $1\%$.

\section{Asymptotically anti-de Sitter black holes}

Let us now turn to the case of a negative cosmological constant, $\LA<0$. Since
$\pt/\pt t$ is timelike in the asymptotic region, there are many qualitative
similarities between these solutions and those discussed in the last section.
There are also some important differences, however. The asymptotic expansions
are given by
$$\eqalign{\PH=\;&{1\over2a}\ln\left|Q\over P\right|+{\Pz{D-1}\over r^{D-1}}+
\OO{1\over r^{D-2}},\cr f=\;&{-2\LA r^2\over{(D-1)(D-2)}}+1-{2M\over r^{D-3}}+{
2\left(Q^2+P^2\right)\over{(D-2)(D-3)r^{2D-6}}}\cr&\hbox to55truemm{\hfil}+{8
\LA\Pz{D-1}^2\over{(2D-3)(D-2)^3r^{2D-4}}}+\OO{1\over r^{2D-3}},\cr R=\;&r-{2
\Db1\Pz{D-1}^2\over{(2D-3)(D-2)^2r^{2D-3}}}+\OO{1\over r^{2D-2}},\cr}\eqn\assia
$$
in terms of the radial coordinate $r$, where we have once again chosen to set
$R\Z0=0$ and to normalise $\Pz0$ according to \setphiz, or by
$$\eqalign{\PH=\;&{1\over2a}\ln\left|Q\over P\right|+{\Pz{D-1}\over R^{D-1}}+
\OO{1\over R^{D-2}},\cr f=\;&{-2\LA R^2\over{(D-1)(D-2)}}+1-{2M\over R^{D-3}}+{
2\left(Q^2+P^2\right)\over{(D-2)(D-3)R^{2D-6}}}\cr&\hbox to55truemm{\hfil}-{8
\LA\Pz{D-1}^2\over{\Db2^3R^{2D-4}}}+\OO{1\over R^{2D-3}},\cr}\eqn\assib$$
in terms of the radial coordinate $R$. In four dimensions the dilaton is
effectively short-range, with $\ph\sim\Pz3r^{-3}$ at spatial infinity, rather
than long-range as for the asymptotically flat solutions. Indeed, this bears
some resemblance to the solutions with a massive dilaton [\GH,\HH], although
there the dependence is one power of $r$ weaker.

The scalar charge, $\Pz{D-1}$, is once again constrained to depend on the other
charges of the theory, and if we integrate \horz\ between the outermost
horizon, $r_+$, and infinity we obtain
$$\Pz{D-1}={a\Db2^2|QP|\over2\LA}\int_{r_+}^\infty\dd r\,{\sinh(2a\PH)\over R^
{D-2}}\,.\eqn\intrelB$$
which by an argument similar to that after \intrelA\ is seen to be negative for
electrically dominated solutions ($|Q|>|P|$), and positive for magnetically
dominated solutions ($|Q|<|P|$).

The next significant difference we have found is that there are no longer
universal discrete eigenvalues of $\g0$ which yield black hole solutions with
two horizons in four dimensions. As in the last section, it is possible
numerically to find values of $\g0$ for which the function $s(R)$ defined by
\shoot\ has zeros at both zeros of $h$. However, these values now always depend
on the initial data and change if one varies any of the parameters $R_-$, $|Q P
|$ or $\LA$. Thus the constraint for obtaining solutions with two regular
horizons would appear to involve $M$, $|QP|$ and $\LA$ in addition to $\g0$, in
some complicated non-linear fashion. Nevertheless, for the wide range of
initial values we have studied we find that the critical values of $\g0$
obtained are fairly close to, but always slightly greater than the
corresponding eigenvalues found for the $\LA=0$ solutions.
\ifig\adsiso{Contours of constant temperature, $T$, for asymptotically anti-de
Sitter black holes (with $\LA=-1$) in four dimensions, as obtained by numerical
integration in the case $\g0=1$: (a) $|P|=0.5|Q|$; (b) $|P|=2|Q|$.}
{\epsfbox{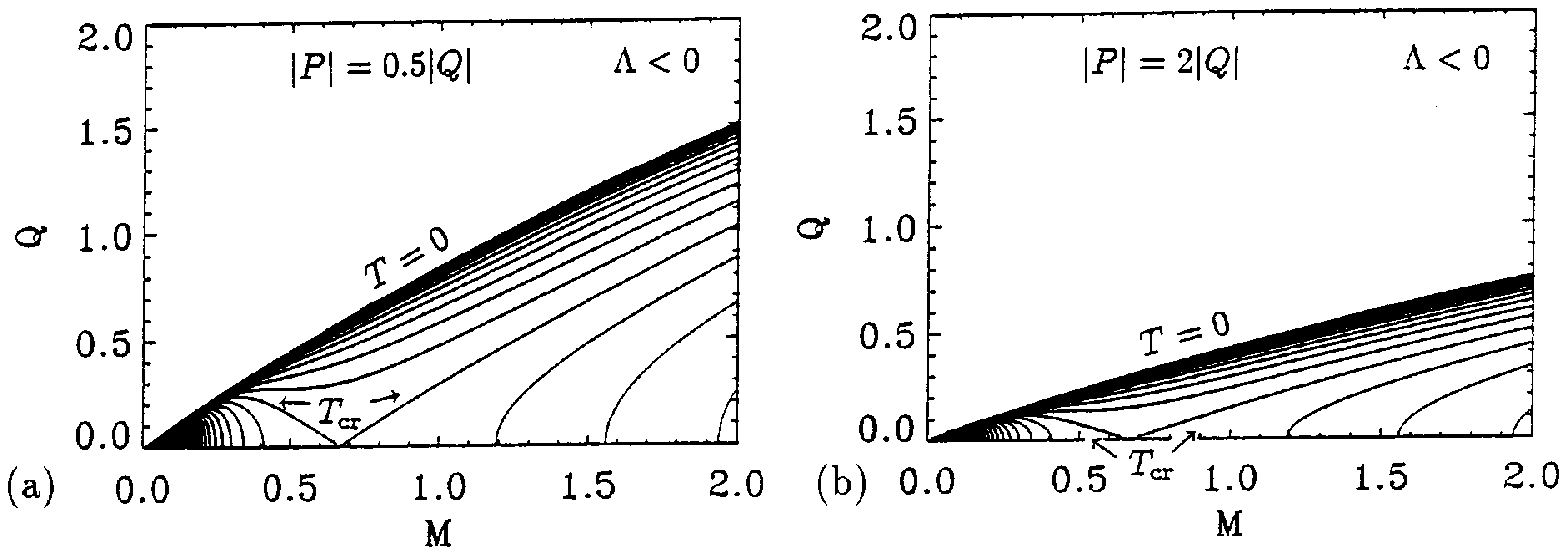}} \ifig\rnadsiso{Contours of constant temperature, $T$, for
the exact \RN-anti-de Sitter black hole solution \rnds\ with $\LA=-1$ and equal
charges $|Q|=|P|$ in four dimensions.} {\epsfbox{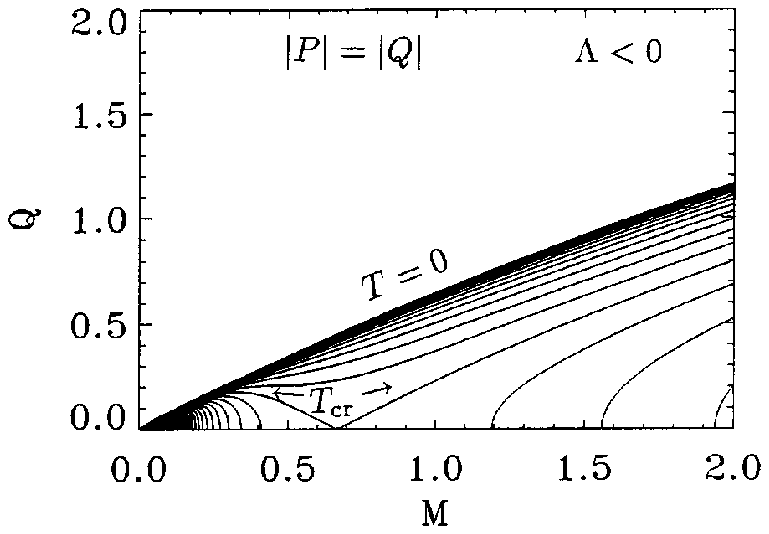}}

Similarly to the case of the asymptotically flat solutions, to observers in the
domain of outer communications the spacetime geometry will appear to be
qualitatively the same as that of the \RN-anti-de Sitter solution -- \rnds\
with $\LA<0$ -- independently of whether the solutions have one or two
horizons. The extreme solutions occur at a finite value of $R$ given by \rbB,
which in four dimensions takes the value
$$R\ns{ext}={-1\over2\LA}\left[\sqrt{1-8\LA|QP|}\,-1\right].\eqn\rbC$$
Once again one can approximate the curve which relates $M$ to $Q$ and $P$ for
the extreme solutions for large values of $M$ and $|QP|$ when the series
\assib\ hold up to the degenerate horizon. However, this is somewhat more
complicated than the asymptotically flat case as it is not possible to obtain a
simple relationship analagous to \relext\ in closed form in this case.

The temperature of the solutions is given by
$$T={1\over4\Db2\pi R_+^{\ 2D-5}}\left[-2\LA R_+^{\ 2D-4}+\Db2\Db3R_+^{\ 2D-6}-
4|QP|\cosh(2a\PH_+)\right]\eqn\tempt$$
where $R_+$ is the outermost horizon and $\PH_+=\PH(R_+)$, and in the extreme
limit $\PH_+=0$, $R_+=R\ns{ext}$ (as given by \rbB) this is zero. We have once
again verified these conclusions by explicit numerical integrations in four
dimensions. The plots shown are for the values $\g0=1$, $\LA=-1$. However,
other values of $\g0\ne0$ and $\LA<0$ lead to results which are qualitatively
the same. The numerically calculated isotherms are shown in Fig.\ 5 for two
values of $|QP|$. For comparison the isotherms of the special case with $|Q|=|P
|$ and a trivial dilaton are shown in Fig.\ 6, as determined from the exact
solutions \rnds. As expected, Figs.\ 5 and 6 are all qualitatively the same,
the only difference being the position of the curve which defines the extreme
limit. By contrast, for asymptotically anti-de Sitter dilaton black holes with
a single charge the extreme limit corresponds to $R\rarr0$ and $T\rarr\infty$
[\PTW], leading to a pattern of isotherms as shown in Fig.\ 7. In all cases the
$Q=0$ axis corresponds to the Schwarzschild-anti-de Sitter solution, which has
a minimum temperature at $T\ns{cr}=\sqrt{-\LA}/(2\pi)$. This critical isotherm
is indicated in the plots. \ifig\isothems{Contours of constant temperature,
$T$, for the singly charged aymptotically anti-de Sitter black hole solution,
as obtained numerically in the case $\g0=1$ (with $\LA=-1$) in four
dimensions.} {\epsfbox{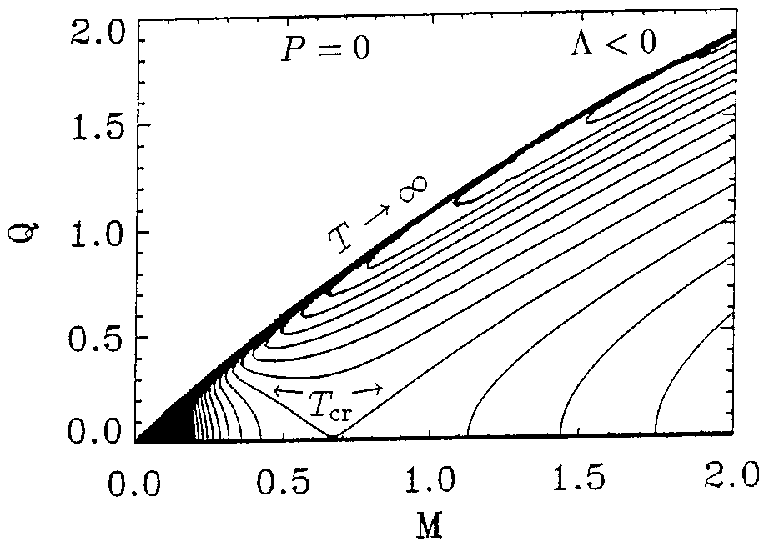}}

\section{Conclusion}

We have shown that the previously known exact solutions [\GM,\Gi--\GW] for
dyonic black holes in four dimensions with a coupling to a non-trivial dilaton
are in fact special cases when one varies the dilaton coupling parameter,
$\g0$. For most values of $\g0$ dyonic solutions have a single horizon, the
exceptions being $\g0\in\{\sqrt{n(n+1)/2},\ n\in\Zop^{\,+}\}$. It is quite
possible that exact solutions could found for the remaining values of $\g0$ in
the series. Despite the absence of an inner horizon, the spacetime geometry is
qualitatively the same as the standard \RN\ solution as far as observers in the
domain of outer communications are concerned, even in the limit of extreme
solutions. The extreme solutions have zero Hawking temperature.

We have further shown that no dyonic black hole solutions with a non-trivial
dilaton exist if $\LA>0$. This is similar to the result obtained for singly
charged dilaton black holes [\PTW], although here there is a solution with a
trivial constant dilaton in the case of equal charges, $|Q|=|P|$.
Asymptotically anti-de Sitter black holes with a non-trivial dilaton do exist
if $\LA<0$, however. These solutions have one or two horizons, the cases of two
horizons being given by a complicated non-linear relationship between $\g0$,
$M$, $|QP|$ and $\LA$. We are unable to determine this relationship
analytically, but we have some numerical evidence for its existence. When
combined with the results concerning singly charged black holes [\PTW] our
results here indicate that the horizon structure of asymptotically anti-de
Sitter dilaton black holes is qualititatively the same as the corresponding
(singly charged or dyonic) asymptotically flat solutions. Once again the
physical properties of the dyonic black holes with $\LA<0$ are the same
regardless of the value of $\g0$, and as in the case of the asymptotically flat
dyonic solutions they have zero Hawking temperature.

Although a pure cosmological constant may not be the most natural cosmological
term in the context of stringy dilaton gravity, we believe that the solutions
we have studied here may nontheless provide a useful approximation in some
circumstances. In particular, if we have a dilaton potential generated by
supersymmetry breaking or some other mechanism, and if the minimum of that
potential which corresponds to the groundstate of the dilaton has a value which
is not precisely zero, then the universe would contain some (hopefully small)
vacuum energy. If this vacuum energy is negative then the black hole solutions
studied here (with $\LA<0$) might provide a useful approximation to the
complete solutions, just as the solutions of [\GH,\HH] are well approximated by
the solutions with a massless dilaton in certain regimes. Phenomenologically a
small positive vacuum energy is currently favoured by astronomers in order to
reconcile the value of the Hubble constant and the age of the universe as
measured by the oldest stars. However, we find no black hole solutions with a
non-trivial dilaton in the case of positive vacuum energy. It is quite
conceivable that this conclusion would be altered in the presence of a
non-trivial dilaton potential, $\V(\ph)$. In particular, the dilaton equation
\feaA\ upon which the non-existence arguments are based -- through \horz\ here
or by an alternative argument for singly charged solutions [\PTW] -- acquires
an additional $\dd\V\over\dd\ph$ term which provides an obstruction to these
arguments for suitable potentials. However, we will leave such considerations
to future work.

\bigskip\noindent{\bf Acknowledgement}\quad This work was supported financially
by the Australian Research Council.

\appendix{A}

We list here the asymptotically flat dyonic black hole solution in arbitrary
spacetime dimension with $\g0=\sqrt{3\Db3}$. The solution is readily generated
from the $D=4$ case \RAB--\EAB, \solB\ if one observes that by using a new
radial coordinate
$$\et=\int R^{D-4}\dd r,\eqn\etta$$
the field equations can be transformed into the $D=4$ equations after a
suitable rescaling of parameters. The solution is
$$\eqalign{R^{D-3}=(D-3)\left[A(\et)B(\et)\right]^{1/4},\cr f=\left(D-3\over R
\right)^{2(D-3)}\left[(\et-M)^2-\et\Z0^{\ 2}\right],\cr\exp\left(4\sqrt{3(D-3)}
\PH\over D-2\right)=\left|Q\over P\right|\left[A(\et)\over B(\et)\right]^{3/2},
\cr}\eqn\crapA$$
where \def\Sii{{2\sqrt{D-3}\,\SI\over\Db2\sqrt{3}}}
$$\eqalign{A(\et)&=\left(\et-\et\Z{A-}\right)\left(\et-\et\Z{A+}\right),\cr B(
\et)&=\left(\et-\et\Z{B-}\right)\left(\et-\et\Z{B+}\right),\cr\et\Z{A\pm}&=-
\Sii\pm\left[8\sqrt{D-3}\,P^2\SI\over\Db2\Db3^3\left[2\sqrt{D-3}\,\SI+\Db2\sqrt
{3}M\right]\right]^{1/2},\cr\et\Z{B\pm}&=\Sii\pm\left[8\sqrt{D-3}\,Q^2\SI\over
\Db2\Db3^3\left[2\sqrt{D-3}\,\SI-\Db2\sqrt{3}M\right]\right]^{1/2},\cr\et\Z0^{\
2}&=M^2+{4\Db3\SI^2\over\Db2^2}-{2\left(Q^2+P^2\right)\over\Db2\Db3^3},\cr
\multispan2\hbox{$\dsp{Q^2\over\sqrt{D-3}\,\SI-\Db2\sqrt{3}M}+{P^2\over\sqrt{D-
3}\,\SI+\Db2\sqrt{3}M}={2\Db3^{7/2}\over\Db2}\,\SI.$}\cr}\eqn\crapB$$
The scalar charge $\SI$ has been normalised so as to correspond to the same
charge in the series \assaf\ when expressed in terms of the radial coordinate
$r$. \refout % \vfil\eject %%for CQG
\figout \end